# Career advancement and scientific performance in universities[1]

*Giovanni Abramo[a], Ciriaco Andrea D'Angelo[a,b], Francesco Rosati[b]*

[a] Institute for System Analysis and Computer Science (IASI-CNR)
National Research Council of Italy

[b] Laboratory for Studies of Research and Technology Transfer
School of Engineering, Department of Management
University of Rome "Tor Vergata"

**Abstract**

Many governments have placed priority on excellence in higher education as part of their policy agendas. Processes for recruitment and career advancement in universities thus have a critical role. The efficiency of faculty selection processes can be evaluated by comparing the subsequent performance of competition winners against that of the losers and the pre-existing staff of equal academic rank. Our study presents an empirical analysis concerning the recruitment procedures for associate professors in the Italian university system. The results of a bibliometric analysis of the hard science areas reveal that new associate professors are on average more productive than the incumbents. However a number of crucial concerns emerge, in particular concerning occurrence of non-winner candidates that are more productive than the winners over the subsequent triennium, and cases of winners that are completely unproductive. Beyond the implications for the Italian case, the analysis offers considerations for all decision-makers regarding the ex post evaluation of the efficiency of the recruitment process and the desirability of providing selection committees with bibliometric indicators in support of evaluation (i.e. informed peer review).

**Keywords**
*Research evaluation; productivity; recruitment; career advancement; competition; bibliometrics; universities; Italy*

# 1. Introduction

The recruitment and the selective advancement of university professors are crucial processes for the medium and long-term future of academic systems. These processes are even more important given the current development of the "world-class" university model, in which one of the essential attributes is concentration of talent (Mohrman et al., 2008; Salmi, 2009). Particularly in "Anglo-Saxon" higher education systems, top universities are in competition with one another to bring in the best teaching professors and researchers from home and abroad. Competitive mechanisms have developed naturally in the Anglo-Saxon nations, with long-standing evolution of national policies that favor the birth and development of a true higher-education market, while the excess of public control seen in many European nations has inhibited the initiation of true competitive mechanisms and led to the development of a generally undifferentiated higher education system that is unable to compete at a global level for access to economic and human resources (public and private funds; talented students, excellent faculty) (Veugelers and van der Ploeg, 2008). According to Auranen and Nieminen (2010), in countries such as Germany, Sweden and Denmark (the present authors would also add Italy), the distinction of different levels of excellence among universities is not immediate, due to an almost total lack of competition among the actors in the system. Yet in many European countries recruitment and advancement take place by rigid procedures, sometimes fixed at the national level, intended - at least in theory - to ensure that competitions select the best candidates for introduction into the system and for gradual stages of advancement. "In theory" should be noted: in the Italian case, for example, recruitment and career advancement do take place by such nationally organized competitions, yet it is far from rare to see cases of favoritism exposed in the press or even arrive before the courts (Zagaria, 2007; Perotti, 2008).

Problems of fairness in selection procedures are certainly not limited to Italy: the international literature has dedicated considerable attention to the study of academic recruitment and promotion, largely regarding questions of gender and minority discrimination (Zinovyeva and Bagues, 2012; van den Brink et al., 2010; Cora-Bramble, 2006; Price et al., 2005; Trotman et al., 2002; Stanley et al., 2007). One of the conclusions is that discriminatory phenomena seem to develop above all when evaluations are based on non-transparent criteria (Rees 2004; Ziegler 2001; Husu 2000; Ledwith and Manfredi 2000; Allen 1988). In effect, academic recruitment is often reported as an informal process in which a few powerful professors select new ones through cooptation mechanisms (van den Brink et al., 2010; Husu, 2000; Fogelberg et al., 1999; Evans, 1995). Such mechanisms often conceal the phenomenon of favoritism, which although it occurs world-wide has been intensively examined in only a few nations, including Turkey (Aydogan, 2012), Australia (Martin, 2009), Spain (Zinovyeva and Bagues, 2012) and Italy (Perotti, 2008). In Italy, there has recently been strong interest in the study of nepotism, which is a particular form of favoritism. While Allesina (2011) and Durante et al., (2011 and 2009) report the unequivocal detection of the phenomenon, Abramo et al. (2012a) are more cautious: they do not deny the presence of nepotism, however they show that the probability a "child" of a parent in the same university not meriting their position is equal to the probability of any "non-child" not deserving their position. This result is in line with what emerges from many sociological studies (Dunn and Holtz-Eakin, 2000; Lentz and Laband, 1989; Simon et al., 1966), suggesting the possibility that "children" employed in the universities may have in fact received a substantial amount of qualifying career-related knowledge from their parents. Whatever the case for nepotism, it does not deal with the concerns for other forms of favoritism that clearly distort faculty recruitment and career advancement, particularly in countries characterized by scarce intensity of competition among universities. Zinovyeva and Bagues (2012), in examining the phenomenon in the Spanish university system, concentrated on the role of connections between the candidates and the evaluators composing the examining boards that decide on academic promotions. They show that the future performance of candidates who were promoted and had a weak connection with the evaluators was better than that of their "non-connected" colleagues. Conversely, successful candidates with a strong link to the evaluators register worse performance both before and after their promotion.

The surveys and case evidence stimulate reflection on the ex-post evaluation of the efficiency of recruitment processes in the academic world. After a certain passage of time from recruitment or advancement in a career it is in fact possible to quantify the "returns" through comparative evaluation of what the academics have achieved in the scientific research sphere[2]. In this work we thus propose to evaluate the efficiency of the selection process for academic career advancement, referring to the Italian university system and in particular to the competitions for associate professor positions (a total of 1,232 posts) held in 2008. We use the bibliometric approach for the evaluation, and for reasons of significance limit the field of observations to the hard sciences. The evaluation is conducted by means of productivity indicators based on scientific publications indexed in the *Thomson Reuters* Web of Science (WoS).

The next section of the paper describes relevant characteristics of the Italian higher education system and the procedures for faculty recruitment, particularly those adopted for the 2008 selections of associate professors. Section 3 illustrates the details of the methodology and dataset used for the analysis. Section 4 presents the results of the analysis in two subsections: first comparing the performance of competition winners and that of associate professors already on staff over the triennium following the 2008 selections; second providing a comparison of the bibliometric productivity of the winners compared to that of the other participants in the competitions, again over the triennium that followed the selections. The work concludes with a summary of the results and the authors' considerations.

## 2. Background

### 2.1 The Italian higher education system

In Italy, the Ministry of Education, Universities and Research (MIUR) recognizes a total of 96 universities as having the authority to issue legally-recognized degrees. Twenty-nine of these are small-sized, private, special-focus universities, of which 13 offer only e-learning. Sixty-seven are public and generally multi-disciplinary universities, scattered throughout the nation. Six of them are *Scuole Superiori* (Schools for Advanced Studies), specifically devoted to highly talented students, with very small faculties and tightly limited enrolment numbers per degree program. In the overall system, 94.9% of faculty are employed in public universities (0.5% in Scuole Superiori). Public universities are largely financed by government through non-competitive allocation. Up to 2009, the core government funding (56% of universities' total income) was input oriented, i.e. independent of merit, and distributed to universities in a manner intended to equally satisfy the needs of each and all, in function of their size and disciplines of research. It was only following the first national research evaluation exercise (VTR), conducted between 2004 and 2006, that a minimal share, equivalent to 3.9% of total income, was attributed by the MIUR in function of the assessment of research and teaching quality.

---

[2] The comparative evaluation of teaching activity is generally more difficult, although we can assume a significant correlation between teaching and research, particularly since quality in teaching will in part follow from scientific merit.

Despite some intervention intended to grant increased autonomy and responsibilities to the universities (Law 168 of 1989[3]), the Italian higher education system is a long-standing, classic example of a public and highly centralized governance structure, with low levels of autonomy at the university level and a very strong role played by the central state.

In keeping with the Humboldtian model, there are no "teaching-only" universities in Italy, as all professors are required to carry out both research and teaching. Time devoted to teaching is established by law: each faculty member must allocate a minimum of 350 hours per year to this area. At the close of 2012, there were 57,400 faculty members in Italy (full, associate and assistant professors) and a roughly equal number of technical-administrative staff. All new personnel enter the university system through public examinations, and career advancement can only proceed by further public examinations, as indicated in the next section.

Salaries are regulated at the centralized level and are calculated according to role (administrative, technical, or professorial), rank within role (for example: assistant, associate or full professor) and seniority. None of a professor's salary depends on merit. Moreover, as in all Italian public administration, dismissal of an employee for lack of productivity is unheard of.

The whole of these conditions have created an environment and culture that are completely non-competitive, yet flourishing with favoritism and other opportunistic behaviors that are dysfunctional to the social and economic roles of the higher education system. The overall result is a system of universities that are almost completely undifferentiated for quality and prestige, with the exception of the tiny Scuole Superiori and a very small number of private, special-focus universities. The system is thus unable to attract significant foreign faculty (or students). The numbers are negligible: only 1.8% of research staff are foreign nationals. This is a system where every university has some share of top scientists, flanked by another share of absolute non-producers. Over the 2004-2008 period, 6,640 (16.8%) of the 39,512 hard-sciences professors did not publish any scientific articles in the journals indexed by the WoS. Another 3,070 professors (7.8%) did achieve publication, but their work was never cited (Abramo et al., 2013a). In bibliometric terms, it means that 9,710 individuals (24.6%) had no impact on scientific progress, as measurable through bibliometric databases[4]. An almost equal 23% of professors alone produced 77% of the overall Italian scientific advancement. This 23% of "top" faculty is not concentrated in a limited number of universities, but is instead dispersed more or less uniformly among all Italian universities, along with the unproductive ones, so that no single institution reaches the critical mass of excellence necessary to develop as an elite university and compete at the international level (Abramo et al., 2012b).

---

[3] This law was intended to grant increased autonomy and responsibility to the universities to establish their own organizational frameworks, including charters and regulations. Subsequently, Law 537 (Article 5) of 1993 and Decree 168 of 1996 provided further changes intended to increase university involvement in overall decision-making on use of resources, and to encourage individual institutions to operate on the market and reach their own economic and financial equilibrium.

[4] Researchers that we define "unproductive" may actually publish in international journals not indexed by Web of Science, or codify the new knowledge produced in different forms, such as books, patents, etc.

**2.2 Recruitment and career advancement**

In Italy, the recruitment and career advancement of professors are regulated by specific law, with its implementation being the responsibility of the MIUR. There have been major reforms of the norms over recent years, with the last one brought about by Law 240 of 2010, which introduced a double level of evaluation for selection of associate and full professors. The first level is national, managed directly by the Ministry, intended to indicate all those candidates with sufficient qualifications in terms of the scientific activity they have conducted; the second is managed by the individual universities and is for the selection of those that are best suited to the specific needs of the university, from among those first judged sufficient at the national level. Prior to Law 240, the processes of recruitment and career advancement were completely in the hands of the individual universities, and in fact the new selection system is still in the start-up phase.

The ex-post analysis of the results of the 2008 competitions is clearly of interest, because not only was this the year the last of the era of exclusively "local" competitions, but also because recruitment procedures had been completely halted for the preceding full three years. For this reason there were competitions launched for thousands of positions, and as these were completed over the next years, the arrivals and advancements of a mass of professors equal to 10% of the total national faculty (12.8% solely among associate professors).

Up until and including 2008, universities would recruit the research and teaching faculty to fill their needs through individual competitions, following the procedures dictated at the central level. These required identification of a committee to judge the curricula of the competition participants and thus choose the best candidates to fill the open position. Each committee was to be composed of five full professors, of which one was designated by the university holding the competition and the other four were chosen from a short list of other full professors in the discipline concerned. The short list was in turn established by national voting among all the full professors of the discipline. The candidate evaluations carried out by each committee were to be based on:
- examination of the documented qualifications presented by each candidate,
- results of an interview held to better understand the candidate's career profile.

For the recruitment of associate professors, the evaluation also included a test of their teaching skill to identify their suitability as an educator in the disciplinary sector.

The law required that the selection committees evaluate the candidates' research activity on the basis of specific criteria, including: innovation, originality and methodological validity of the scientific production; contribution to work completed in collaboration with others; conformity of the candidate's research history with the scientific discipline indicated by the competition; scientific relevance of the journals or other media for their publications, and their resulting diffusion within the scientific community; timeliness of the scientific production in relation to the evolution of knowledge in the disciplinary sector. The personal documentation to be evaluated was to concern: history of teaching activity; employment service in national and foreign universities and research institutes; organization, direction and coordination of research groups and/or initiatives in teaching and research.

After individual and joint judgments of all the candidates, the committee members as a group were required to vote for selection of two winning candidates[5]. At that point the university that held the competition was free to hire one of the two winners for the announced position while the other remained eligible for hiring over the next five years, in analogous rank and without further competition, by any other university in the national system.

In order to rationalize the process of the individual competitions over the entire system, the MIUR monitored and gathered the hiring proposals of the various universities and supported the evaluation procedures through information management aimed at better guaranteeing transparency. One of the ministry measures was to provide a Web portal with all the basic information on the competition procedures, posts available, numbers of candidates for each competition, the scheduling of the procedures and final results.

The transparency provisions, nomination of a national committee of experts in the sector, and the timely issue of regulations for the evaluation procedures were all intended to ensure efficient processes and the achievement of the desired objectives. In reality, the characteristics of Italian system - such as the generally strong inclination to favoritism, the structured absence of responsibility for poor performance by research units, and the lack of incentive schemes for merit - undermined the credibility of selection procedures for hiring and advancement of university personnel, just as happens for the public administration in general. This is demonstrated by the high and growing number of legal cases brought by losing candidates and by specific studies of the systemic problems (Perotti, 2008; Zagaria, 2007) - and last but not least by the unavoidable evidence of the 25% of professors that are unproductive.

Now that some years have passed since the 2008 event of 1,232 competitions for career advancement, we take the notable opportunity of evaluating the efficiency of the selection process, relative to the results of scientific activity achieved over the subsequent triennium.

3. Methodology

3.1 Measuring research performance at individual level

Research activity is a production process in which the inputs consist of human, tangible (scientific instruments, materials, etc.) and intangible (accumulated knowledge, social networks, etc.) resources, and where outputs have a complex character of both tangible nature (publications, patents, conference presentations, databases, protocols, etc.) and intangible nature (tacit knowledge, consulting activity, etc.). The new-knowledge production function has therefore a multi-input and multi-output character.

---

[5] The committees could also indicate a single winner, however in reality this occurred very rarely.

The principal efficiency indicator of any production system is labor productivity. To calculate it one needs to adopt a few simplifications and assumptions.

It has been shown (Moed, 2005) that in the hard sciences the prevalent form of codification of research output is publication in scientific journals. As a proxy of total output, in this work we consider only publications (articles, article reviews, and proceeding papers) indexed in the WoS. The other forms of output which we neglect are often followed by publications that describe their content in the scientific arena, so the analysis of publications alone actually avoids a potential double counting.

When measuring labor productivity, if there are differences in the production factors available to each scientist then one should normalize by them. Unfortunately relevant data are not available at individual level in Italy. The first assumption then is that the resources available to professors within the same field of observation are the same. The second assumption is that the hours devoted to research are more or less the same for all professors. Given the main traits of the Italian academic system as depicted in section 2.1, the above assumptions appear acceptable.

Because of the differences in the publication intensity across fields, a prerequisite of any research performance assessment free of distortions is a classification of each researcher in one and only one field (Abramo et al., 2013b). In the Italian university system all professors are classified in one field (named scientific disciplinary sectors - SDSs, 370 in all), grouped into disciplines (named university disciplinary areas - UDAs, 14 in all)[6]. To our knowledge, this feature of the Italian higher education system is unique in the world. Furthermore, since it has been demonstrated that productivity of full, associate and assistant professors is different (while their distribution is not uniform across universities, and academic rank determines differentiation in salaries) in order to avoid distortions in performance rankings (Abramo et al., 2010a), we differentiate them by academic rank.

A very gross way to calculate the average yearly labor research productivity of a scientist is to simply count publications (O, output) in the period of observation and divide it by the full-time equivalent of work in the period. A more sophisticated way recognizes that publications, embedding the new knowledge produced, have different values that bibliometricians approximate with citations. In very specific cases, when the time window between citations count and publication date is short (two years or shorter), it has been shown that IF is a better proxy of expected impact than citations (Abramo et al., 2010b). In all other cases, the use of citations must be always preferred (Abramo et al., 2011). In our specific case, because we measure productivity in the 2009-2011 period and the available citations were observed on 31/12/2011, we had to resort to the impact factor of journals to proxy the value of publications.

Since citation behavior and consequently impact factors vary across fields, and it has been shown (Abramo and D'Angelo, 2011) that it is not unlikely that researchers belonging to a particular scientific field may also publish outside that field[7], we must standardize impact values with respect to a scaling factor pertaining to the distribution for the same year and subject category[8]. Because of the constraint of the very short citation window, for this specific study, we substitute citations with the impact factor in the formula of the productivity indicator, Fractional Scientific Strength, *FSS*, that we normally use (Abramo et al., 2013b). To highlight the difference, we add a subscript *IF*:

$$FSS_{IF} = \frac{1}{t}\sum_{i=1}^{N}\frac{IF_i}{\overline{IF_i}}f_i$$

---

[6] The complete list is accessible on http://attiministeriali.miur.it/UserFiles/115.htm, last accessed on March 4, 2013.

[7] A typical example is statisticians, who may apply their theory to medicine, physics, social sciences, etc.

[8] For details, see Abramo et al., 2013b.

Where:

$IF_i$ = journal impact factor of publication *i*;
$\overline{IF_i}$ = average impact factor of all journals in the same subject category of publication *i*;
$f_i$ = fractional contribution of the researcher to the publication i. Fractional contribution equals the inverse of the number of authors, in those fields where authors appear in alphabetical order in the byline, but assumes different weights in the life sciences, where we give different weights to each co-author according to their order in the byline and the character of the co-authorship (intra-mural or extra-mural)[9] (see Abramo et al., 2013c);
N = number of publications of the researcher in the period of observation;
t = number of years of work of the researcher in the period of observation.

**3.2 Dataset**

Data on faculty of each university and their SDS classification are extracted from the database on Italian university personnel maintained by the MIUR[10]. The bibliometric dataset used to measure O and $FSS_{IF}$ is extracted from the Italian Observatory of Public Research (ORP), a database developed and maintained by the authors, derived under license from the WoS. Beginning from the raw data of the WoS, and applying a complex algorithm for reconciliation of the author's affiliation and disambiguation of the true identity of the authors, each publication (article, article review and conference proceeding) is attributed to the university scientist or scientists that produced it (D'Angelo et al., 2011). Thanks to this algorithm we can produce rankings of research productivity at the individual level, on a national scale. Based on the value of O and $FSS_{IF}$ we obtain, for each SDS, a ranking list expressed in percentiles and differentiated by academic rank. Thus the performance of each scientist is expressed on a scale of 0-100 (worst to best), comparing to the performance of all Italian colleagues of the same SDS and academic rank. In the current study we use both indicators of performance: the intermediate, inaccurate O and the more sophisticated and accurate $FSS_{IF}$, embedding both the standardized impact of output and the fractional contribution of authors. In this manner we can search for evidence of the potential difficulties and shortcomings of the committees in carrying out efficient selections - in their case without support from bibliometric measures (informed peer review).

---

[9] If first and last authors belong to the same university, 40% of the contribution is assigned to each of them; the remaining 20% is divided among all other authors. If the first two and last two authors belong to different universities, 30% of the contribution is assigned to first and last authors; 15% is attributed to second and last author but one; the remaining 10% is divided among all others. The weighting values were assigned following advice from senior Italian professors in the life sciences. The values could be changed to suit different practices in other national contexts.
[10] http://cercauniversita.cineca.it, last accessed on March 4, 2013.

Overall in 2008, 1,232 competitions for associate professor positions[11] were launched by a total of 74 universities. Table 1 provides the details per individual UDA. In total the competitions concerned 299 SDSs. At the end of all the processes, which lasted an average of over two years[12], the committees had named 2,339 winners, out of a total of 16,500 candidates. The comparison between the last two columns of the table clearly shows the impact of the 2008 round of competitions. The ratio of number of winners of competitions to the size of the existing teaching Italian faculty averages 12.8%, varying from a minimum of 8.7% in Earth sciences to a maximum of 21% in Law. In five UDAs the applications received were numerically higher than the number of permanent faculty at the time. For Industrial and information engineering, compared to 1,493 associate professors on faculty, there were 2,010 applications for 144 competitions launched by 31 universities.

*Table 1: Data on year 2008 competitions for associate professor posts, per UDA*

| UDA | Compet. launched | SDSs concerned | Univ. launching compet. | Applications submitted | Winners* | Associate prof. on staff 2008 |
|---|---|---|---|---|---|---|
| Mathematics and computer science | 59 | 9 out of 10 | 26 | 1,404 | 114 | 1,092 |
| Physics | 46 | 6 out of 8 | 24 | 1,102 | 90 | 891 |
| Chemistry | 55 | 11 out of 12 | 24 | 766 | 104 | 1,040 |
| Earth sciences | 20 | 11 out of 12 | 12 | 268 | 37 | 423 |
| Biology | 75 | 18 out of 19 | 33 | 1,387 | 147 | 1,517 |
| Medicine | 168 | 44 out of 50 | 35 | 2,338 | 316 | 3,298 |
| Agricultural and veterinary sciences | 56 | 24 out of 30 | 21 | 460 | 107 | 939 |
| Civil engineering and architecture | 86 | 20 out of 22 | 30 | 1,291 | 161 | 1,201 |
| Industrial and information engineering | 144 | 36 out of 42 | 31 | 2,010 | 279 | 1,493 |
| Ancient history, philology, literature and art history | 103 | 41 out of 77 | 32 | 943 | 192 | 1,794 |
| History, philosophy, pedagogy and psychology | 118 | 30 out of 34 | 42 | 1249 | 220 | 1,547 |
| Law | 128 | 21 out of 21 | 49 | 1148 | 245 | 1,169 |
| Economics and statistics | 133 | 17 out of 19 | 43 | 1720 | 250 | 1,341 |
| Political and social sciences | 41 | 11 out of 14 | 23 | 483 | 77 | 512 |
| Total | 1,232 | 299 out of 370 | 74 | 16,569 | 2,339 | 18,257 |

*\* The data indicate the results declared from the 1,221 competition procedures that were officially completed (out of 1,232 launched) at the time of preparing the current research paper.*

---

[11] Retrieved from: http://reclutamento.murst.it/, the open Web site managed by the MIUR, titled "Comparative evaluation in the recruitment of University Professors and Researchers (Law 3, 3 July 1998, no. 210)".

[12] At the time of the current research, eleven competitions had not been completed.

To ensure the representativity of publications as proxy of the research output, our analysis is focused on the competitions concerning SDSs where at least 50% of professors produced at least one publication indexed in the WoS over the period 2004-2008. With these restrictions, the analysis is limited to 654 competitions concerning 193 SDSs (185 belonged to the nine UDAs listed in Table 1; we arbitrarily group the remaining eight[13] as the generic UDA "Other"). All of the competitions but 39 announced two winners. Of the total 1,269 winners, 91.3% (1,159) were academics already on staff as assistant professors.

In order to compare the research productivity of the winners with that of the other applicants it was necessary to identify all the applicants, by reading the minutes of each competition (published on-line by the individual universities). Given the prohibitive scope of this task, we selected 287 competitions (44% of the total 654) that were organized by 12 universities: the four institutions with the highest numbers of launched competitions from each of the northern, central and southern national areas. The data for this subset are presented in Table 2. For this subpopulation, the winners (550 in all) represent 22% of the total candidates (2,590); the rate of selection was slightly more favorable for incumbent assistant professors from Italian universities (532/2,314=23.0%) than it was for other candidates (18/276=6.5%). On average there were nine participants per competition, of which eight were Italian-national academics. However the number of candidates shows significant variation (Std Dev. 5.6), with 16 competitions where there were 20 or more candidates and one that reached the level of 29 candidates.

*Table 2: Winners and candidates of 287 competitions examined in detail*

|  | Winners | Non winners | Total candidates | Candidates per competition | | | |
|---|---|---|---|---|---|---|---|
|  |  |  |  | Average | Median | Std Dev. | Max |
| Italian academics | 532 | 1,782 | 2,314 | 8 | 7 | 5.4 | 28 |
| Others | 18 | 258 | 276 | 1 | 1 | 1.2 | 6 |
| Total | 550 | 2,040 | 2,590 | 9 | 8 | 5.6 | 29 |

## 4. Results and discussion

In this section we evaluate the scientific productivity of the winners of the 2008 competitions over the 2009-2011 triennium, in comparison to performance by:
- associate professors on staff in 2008;
- other applicants to the competitions, on staff as assistant professors.

The decision to evaluate performance over the post-competition triennium arises from the general interest in analyzing the capacity of nominated committees to evaluate, "on the ground", the best candidates in terms of their potential future scientific productivity. By comparing the performance of the winners against that of the incumbent associate professors we intend to identify if the winners effectively reach levels of productivity that are adequate to the academic ranks for which they were chosen. By comparing between the performance of the winners and that of the "non-winners", our intention is to understand if the committees' personnel choices have proven correct in terms of the expected scientific results. In the analysis we use both of the productivity indicators (O and $FSS_{IF}$) as defined above. For both indicators, each scientist's performance is expressed on a scale of 0-100 (worst to best), comparing to the performance of all Italian colleagues of the same SDS and academic rank. Because of the limits of WoS coverage and the approximations embedded in bibliometrics indicators and methods, our findings need to be interpreted with special caution.

---

[13] M-PSI/01: General Psychology; M-PSI/02: Psychobiology and Physiological Psychology; M-PSI/03: Psychometrics; M-EDF/01: Teaching Methods for Physical Activities; M-EDF/02: Teaching Methods for Sport; SECS-P/05: Econometrics; SECS-S/01: Statistics; SECS-S/06: Mathematics for Economics, Actuarial Studies and Finance.

### 4.1 Research performance of winners as compared to that of incumbent associate professors

To ensure greater significance for this first analysis we consider only professors that were on staff for all three years from to 2009 to 2011 (Table 3).

Considering the output indicator (O), the average percentile rank of the winners is 67.1, compared to 49.1 for the incumbent associate professors. For fractional scientific strength ($FSS_{IF}$), the average percentile is 63.5 for winners and 46.1 for incumbents. The Student's *t*-test for significance in the differences gave positive results (two tailed *P*-value < 0.01). The share of winners with no publications is 4.5% of total, compared to 16.0% of the incumbents; the share of winners with publications in journals with no impact ($FSS_{IF} = 0$) is 5.5%, compared to 18.2% of the incumbents. Considering $FSS_{IF}$, 8.3% of the winners place in the bottom 20% of scientists in each SDS and 28.9% have performance below the median, compared to 26.0% and 52.2% for the incumbents. 34.6% of the winners place in the top 20% scientists, compared to 18.5% of the incumbents. Raising the threshold to the top 10% of scientists, we see that 19.2% of the winners belong to the "excellent" category, compared to only 9.4% of incumbents.

*Table 3: Comparison of 2009-2011 research performance ($FSS_{IF}$) by winners and incumbents*

|  | Winners | Incumbents |
|---|---|---|
| Observations | 1,154* | 9,297 |
| Average percentile rank for O | 67.1 | 49.1 |
| Average percentile rank for $FSS_{IF}$ | 63.5 | 46.1 |
| Professors with no publications (%) | 4.5 | 16.0 |
| Professors with no impact (%) | 5.5 | 18.2 |
| Bottom 20% scientists for $FSS_{IF}$ (%) | 8.3 | 26.0 |
| Below median for $FSS_{IF}$ (%) | 28.9 | 52.2 |
| Top 20% scientists for $FSS_{IF}$ (%) | 34.6 | 18.5 |
| Top 10% scientists for $FSS_{IF}$ (%) | 19.2 | 9.4 |

*\* The difference from the total set of 1,159 academic winners (Section 3.2) is due to five assistant professors who won two different competitions: here they are counted only once.*

The comparison between research performance ($FSS_{IF}$) of winners and of incumbents broken down by UDA (Table 4) again shows significant differences according to Student's *t*-test (all two tailed p-value < 0.01). The most notable difference in productivity between winners and incumbents is in the Mathematics and computer science UDA: for the winners the average percentile rank is 69.3, while for the incumbents it is 42.7. The difference is also notable in the Earth sciences UDA: for the winners the average percentile rank is 68.6, compared to 45.6 for incumbents. The UDA showing the least difference is Biology: for the winners the average percentile rank equals 60.9 and for the incumbents it equals 48.3.

The Civil engineering and architecture UDA is the one with the highest share of winning professors with no publications (14.0%), and with no publications in journals with impact factor (16.0%)[14]. This is also a UDA where the incumbent population as a whole has very high non-productive shares (respectively 32.3% and 36.9%). Biology and Chemistry are the two UDAs with the highest share of winners with $FSS_{IF}$ below the median: 34.1% for Biology and 33.7% for Chemistry.

*Table 4: Comparison of 2009-2011 research performance ($FSS_{IF}$) of winners (W) and incumbents (I) by UDA*

| UDA | Observations | | Average percentile rank for $FSS_{IF}$ | | Professors with no publications (%) | | Professors with no impact factor (%) | | Below Median for $FSS_{IF}$ (%) | | Top 20% scientists for $FSS_{IF}$ (%) | |
|---|---|---|---|---|---|---|---|---|---|---|---|---|
| | W | I | W | I | W | I | W | I | W | I | W | I |
| Agricultural and veterinary sciences | 90 | 610 | 63.1 | 45.8 | 3.3 | 15.7 | 3.3 | 16.7 | 28.9 | 52.5 | 33.3 | 18.7 |
| Biology | 132 | 1,262 | 60.9 | 48.3 | 2.3 | 9.7 | 3.0 | 9.9 | 34.1 | 51.3 | 34.8 | 18.8 |
| Chemistry | 95 | 880 | 64.4 | 48.2 | 0.0 | 6.3 | 0.0 | 6.5 | 33.7 | 51.5 | 35.8 | 18.5 |
| Civil engineering and architecture | 50 | 347 | 62.8 | 41.2 | 14.0 | 32.3 | 16.0 | 36.9 | 24.0 | 53.9 | 30.0 | 19.0 |
| Earth sciences | 31 | 294 | 68.6 | 45.6 | 3.2 | 19.4 | 3.2 | 21.4 | 22.6 | 52.0 | 41.9 | 18.4 |
| Industrial and information engineering | 264 | 1,287 | 63.2 | 44.2 | 4.2 | 16.3 | 6.1 | 23.6 | 29.2 | 53.8 | 33.3 | 17.6 |
| Mathematics and computer science | 108 | 890 | 69.3 | 42.7 | 3.7 | 26.0 | 8.3 | 31.2 | 22.2 | 53.1 | 46.3 | 17.0 |
| Medicine | 239 | 2,601 | 63.2 | 47.3 | 5.4 | 15.1 | 5.4 | 15.7 | 27.2 | 51.7 | 33.1 | 19.1 |
| Physics | 77 | 724 | 62.7 | 48.2 | 1.3 | 9.5 | 1.3 | 10.1 | 32.5 | 51.7 | 31.2 | 18.9 |
| Other | 68 | 402 | 60.1 | 40.9 | 13.2 | 35.6 | 13.2 | 38.1 | 30.9 | 52.5 | 29.4 | 18.9 |

The data presented above clearly show that in the triennium following the competitions, the winners performed better than their colleagues already on staff, in terms of scientific productivity. However the average values do not disguise cases of inadequacy among the winners: among these, 4.5% did not produce any publications from 2009 to 2011 and a further 1% published only in journals with no impact factor. Given these situations of unexpected results from certain competitions, we conducted a further analysis. Table 5 shows the number of competitions (absolute and as % of total) where the winners were unproductive (output=0; or $FSS_{IF}$=0), placed in the bottom 20%, or placed below the median. Research performance is again measured with the two productivity indicators O and $FSS_{IF}$ over the 2009-2011 period. For reasons of significance we analyze only the competitions were both winners were on staff over all three years from 2009 to 2011, for a total of 513 observations out of the starting 654.

For 8.4% of the competitions (43 out of 513), at least one winner produced no publications over the 2009-2011 period; in a further 2.3% of cases (12 of the competitions) at least one winner did publish but only in journals with no impact factor.

---
[14] In this UDA WoS coverage is quite limited, which shows in a very high number of non-productive (in bibliometric terms) professors. In this UDA, our findings need to be interpreted with additional special caution.

The competitions where at least one winner placed in the bottom 20% number 60 of total 513 (11.7%) as measured by indicator O, and 80 (15.6%) on the basis of $FSS_{IF}$. In 39.8% of the competitions (204 out of 513), at least one winner registers O below the median and in 48.1% of the competitions (247 su 513) at least one winner shows $FSS_{IF}$ below the median. Particularly striking cases are those where both of the competition winners went on to produce no publications over the next three years and three competitions where both winners register nil $FSS_{IF}$.

*Table 5: Distribution of 2008 competition winners for scientific productivity over the 2009-2011 period*

| Competitions | O | | $FSS_{IF}$ | |
|---|---|---|---|---|
| | At least one | Both | At least one | Both |
| With unproductive winners | 43 of 513 (8.4%) | 1 of 513 (0.2%) | 55 of 513 (10.7%) | 3 of 513 (0.6%) |
| With winners in bottom 20% | 60 of 513 (11.7%) | 2 of 513 (0.4%) | 80 of 513 (15.6%) | 3 of 513 (0.6%) |
| With winners below the median | 204 of 513 (39.8%) | 26 of 513 (5.1%) | 247 of 513 (48.1%) | 48 of 513 (9.4%) |

In terms of output, the average performance of the two winners is below the median in 86 competitions out of 513, or 16.8% of total (Table 6); considering $FSS_{IF}$, we see the same characteristic in 126 out of 513 competitions (one quarter of total). Another, certainly interesting situation is that where performance of one winner is below median, while at the same time the other performs in the top 20%. This occurs in 14.8% of cases (76 competitions out of 513) for indicator O, and in 90 competitions (17.5% of cases) for $FSS_{IF}$.

*Table 6: Analysis of 2008 competitions on the basis of performance by the winners, 2009-2011*

| Competitions | O | $FSS_{IF}$ |
|---|---|---|
| Where average performance of the winners is below the median | 86 of 513 (16.8%) | 126 of 513 (24.6%) |
| With one winner in bottom 20% and the other top 20% | 21 of 513 (4.1%) | 23 of 513 (4.5%) |
| With one winner below median the other in top 20% | 76 of 513 (14.8%) | 90 of 513 (17.5%) |

**4.2 Research performance of winners as compared to that of other candidates**

The analyses in the preceding section show that over the triennium considered, the scientific activity of competition winners was an average greater than that of colleagues who were already on staff as associate professors, however there is no lack of cases where the productivity of winners falls below. We could imagine that in some competitions all the candidates had a low scientific profile and the committees, having in any case to proceed, would use comparative evaluation to choose the "best from the bad". We return to this potential scenario in the conclusions to our paper; however, whatever the range of candidates in the competitions, we wish to compare the scientific productivity of the winners with that of other candidates, to understand the frequency of cases where committees made an inefficient selection. To do this it was first necessary to select a sample of competitions, given that the identification of the true candidates can only be accomplished by the labor-intensive scrutiny of all the individual

Committee minutes, to reveal those candidates who were actually subject to evaluation, from among all those who initially applied[15]. For this analysis we thus refer to the subset of 287 competitions indicated in Table 2, meaning those launched by the four northern, central and southern universities that were most active in conducting competitions. Of these 287 competitions, there are 205 with at least one "non-winner" participant that can be considered in our evaluation[16], and 196 with at least two non-winner participants.

Table 7 presents the number of competitions where at least one or two non-winner participants had O or $FSS_{IF}$ greater than one or both the winners, over the 2009-2011 period.

There are 178 out of 204 (87.3%) competitions where at least one non-winner achieved greater output than at least one winner over the subsequent triennium. Considering $FSS_{IF}$, the number of such competitions rises to 182 (89.2%). In 74 competitions of 196 (37.8%), at least two non-winners achieved greater output than both winners. For $FSS_{IF}$, the number of competitions rises to 75 (38.3%).

In 167 competitions out of 204 (81.9%) there was at least one non-winner participant who was more productive than a winner, for both indicators, and in 131 competitions of 196 (66.8%) at least two non-winners achieved higher productivity than a winner, again for both indicators. In 83 of 204 competitions (40.7%) at least one non-winner obtained a rank for both indicators higher than that of both the winners; in 50 competitions of 196 (25.5%), there were at least two non-winners with both indicators of performance higher than those of both winners.

*Table 7: Number (percentage) of competitions in which at least one or two participants had performance greater than one or both winners*

| Competitions | O | $FSS_{IF}$ | Both O and $FSS_{IF}$ |
|---|---|---|---|
| Where at least one participant had performance greater than at least one winner | 178 of 204 (87.3%) | 182 of 204 (89.2%) | 167 of 204 (81.9%) |
| Where at least two participants had performance greater than at least one winner | 140 of 196 (71.4%) | 153 of 196 (78.1%) | 131 of 196 (66.8%) |
| Where at least one participant had performance greater than both winners | 111 of 204 (54.4%) | 120 of 204 (58.8%) | 83 of 204 (40.7%) |
| Where at least two participants had performance greater than both winners | 74 of 196 (37.8%) | 75 of 196 (38.3%) | 50 of 196 (25.5%) |

Figure 1 provides further detail concerning Table 7, showing the number of competitions where 1, 2, … *n* non-winner candidates outperformed one or both winners. The analysis reveals many situations of inefficient selection, in other words competitions where a substantial number of participants went on to achieve greater scientific productivity than the winners. For example, we observe 22 competitions (of 204 analyzed) where not less than 10 participants register productivity higher than that of one of the winners (last light-colored bar of the histogram), or the case of the 30 competitions where not less than five participants register productivity higher than both winners (aggregation of all dark-shaded bars, axis "5" and up). The data are that much more significant if we consider that the number of candidates per competition (while not uniform) is on average quite low: as indicated in Table 2, there were an average of nine applicants per competition and of these there were an average of only 8 (academic) candidates that could be included in our analysis.

---

[15] The number of candidates that actually arrived at the stage of committee evaluation was consistently less than the initial applicants, especially in SDSs where many universities launched competitions.

[16] For our own evaluation, there must be at least one non-winner who held the role of assistant professor over the subsequent triennium, for at least one year.

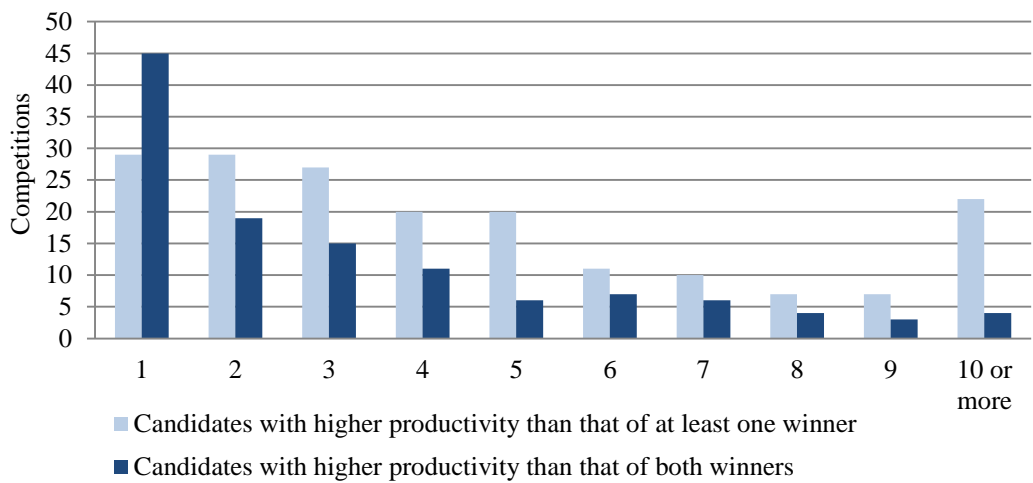

*Figure 1: Number of competitions where "non-winner" participants show higher productivity ($FSS_{IF}$) than winners*

Even though the bibliometric methodology adopted would have evident limits in simulating the complex evaluation mechanisms that committees might have implemented, to select the individuals best-suited as associate professors from among all the candidates, the results of the analyses show without doubt that in the triennium following the competitions the non-winners very often had research productivity that was clearly higher than that of winners.

## 5. Conclusions

Recruitment and career advancement are highly critical processes for any productive system: even more so for higher education systems, where many nations are investing substantial resources, in spite of the economic downturn, specifically to relaunch their economies. The efficiency of these processes can be evaluated by comparing the subsequent research performance of the recruitment "winners" with that of competition losers and pre-existing incumbents of equal academic rank. In Italy the last recruitment campaign for university faculty took place in 2008, mostly to fill associate professor positions from the ranks of assistants, with some competitions to rise from associate to full professor. The 2008 competitions gave rise to a massive renewal of faculty, with insertion of new associate professors totaling almost 13% of the staff already holding such academic rank. The current study evaluates the scientific activity of the competition winners over the triennium following the selections.

A first analysis revealed that over the 2009-2011 period, the average performance of the winners was significantly higher than that of their colleagues who were already on staff as associate professors. However it also emerged that 29% of winners had productivity lower than the median in their SDS, and that 5.5% of winners did not achieve any citations indexed in WoS. The analysis of the single competitions shows that almost half of them selected at least one winner that achieved productivity below the median for their field over the following triennium; in 10% of competitions this occurred for both winners.

A deeper analysis for a subset of competitions, in which we identified all the candidates, shows that in two thirds of the cases at least two non-winner participants registered productivity greater than at least one of the winners. In fact in a quarter of the competitions, both winners showed productivity that was systematically lower than at least two non-winner participants. The inefficiencies observed in the selections are still more pronounced when we consider the more refined productivity indicator, $FSS_{IF}$. The results raise questions about the general need to provide university recruitment committees with bibliometric indicators in support of their evaluations (informed peer review).

We note that the 2008 Italian recruitment campaign was conducted under a regulatory framework that recent legislation has completely modified, including providing for new preselection mechanisms intended to reduce potential distortion phenomena inherent in peer evaluations. But it will take many years to understand if the new regulatory framework - which has still to be fully implemented - will solve the critical problems of the preceding one and guarantee efficient selection in the national academic system. The intention of the authors has not been to raise any doubts about ethical issues in the personnel selections made, only to understand to what extent the winners merited the trust of the committees that they would implement one of the fundamental missions of the academic system: knowledge advancement.

Such university selection committees carry out their activity as shareholder delegates, which in the case of prevalently public systems such as in Italy, include all taxpayers. Given the facts of efficiency and issues of trust, our study has revealed several critical concerns that should now be further investigated relative to the mass of faculty inserted in the Italian academic system by the 2008 procedures. In particular, the authors perceive the necessity for further analysis of the determinants of the observed results. It should certainly be of interest to measure the scientific productivity of the evaluators, to attempt correlations with that of the winners. It also seems useful to investigate the presence of proximity effects that might have given advantage to candidates that were socially closer to one or more evaluators. Finally, it would appear useful to stratify the results of the analysis by geographic area, to detect potential regional peculiarities, as well as to repeat the analysis under the new recruitment system once it takes effect, to understand the depth and breadth of the resulting changes in the scientific profiles of the winners.